\documentclass[]{interact}

\usepackage{epstopdf}% To incorporate .eps illustrations using PDFLaTeX, etc.
\usepackage[caption=false]{subfig}% Support for small, `sub' figures and tables

\usepackage[numbers,sort&compress]{natbib}% Citation support using natbib.sty
\bibpunct[, ]{[}{]}{,}{n}{,}{,}% Citation support using natbib.sty
\renewcommand\bibfont{\fontsize{10}{12}\selectfont}% Bibliography support using natbib.sty
\makeatletter% @ becomes a letter
\def\NAT@def@citea{\def\@citea{\NAT@separator}}% Suppress spaces between citations using natbib.sty
\makeatother% @ becomes a symbol again

\theoremstyle{plain}% Theorem-like structures provided by amsthm.sty

\theoremstyle{definition}

\theoremstyle{remark}

\usepackage{listings}
\usepackage{color}
\def\red{\color{black}}
\begin{document}

\articletype{RESEARCH ARTICLE}% Specify the article type or omit as appropriate

\title{CluBear: A Subsampling Package for Interactive Statistical Analysis with Massive Data on A Single Machine}

\author{
\name{Ke Xu\textsuperscript{a}, Yingqiu Zhu\textsuperscript{a}\thanks{CONTACT Yingqiu Zhu. Email: rozen0maiden@126.com}, Yijing Liu\textsuperscript{b} and Hansheng Wang\textsuperscript{c}}
\affil{\textsuperscript{a}School of Statistics, University of International Business and Economics, Beijing, China; \textsuperscript{b}Beijing PERCENT Technology Group Co., Ltd., Beijing, China; \textsuperscript{c}Guanghua School of Management, Peking University, Beijing, China;}
}

\maketitle

\begin{abstract}
This article introduces {\it CluBear}, a Python-based open-source package for interactive massive data analysis. The key feature of {\it CluBear} is that it enables users to conduct convenient and interactive statistical analysis of massive data with only a traditional single-computer system. Thus, {\it CluBear} provides a cost-effective solution when mining large-scale datasets. In addition, the CluBear package integrates many commonly used statistical and graphical tools, which are useful for most commonly encountered data analysis tasks.
\end{abstract}


\section{Introduction}\label{sec:intro}

Massive datasets are becoming increasingly common in modern statistical analyses \citep{rajaraman2011mining,national2013frontiers,li2013statistical,xu2016prediction,lian2019projected,chen2021distributed}. How to perform convenient and efficient statistical analysis of massive datasets has thus become challenging. One feasible solution is to rely on a powerful distributed system that contains many computers \citep{mcdonald2009efficient,zinkevich2010parallelized,YuchenZhang2013CommunicationefficientAF,suresh2017distributed,jordan2018communication,huang2019distributed}. For a distributed system, computers are physically connected to each other and managed via appropriately designed software (e.g., Hadoop, Spark). Distributed computation can then be performed to process massive data efficiently. Using these systems, various techniques for distributed computation have been proposed \citep{YuchenZhang2013CommunicationefficientAF,QiangLiu2014DistributedEI,BenjaminDenham2020HDSMAD,HamidBeigy2021ASM,DaejinKim2021OptimalLA}. Distributed computation provides a useful and promising solution but is typically expensive because a well-maintained distributed system is not easily accessible to an individual user \citep{kane2013scalable}. Most general users must perform standard statistical computations using their personal computers or workstations. These computers are typically equipped with a hard disk that is sufficiently large (e.g., 1 TB) to hold massive datasets. However, their memories  are often limited (e.g., 32 GB). The numbers of CPU cores are also limited. As a result of the insufficiency of memory or CPUs, a computational environment within a single-computer system can typically not support massive data analysis \citep{SaidHamdioui2019TestingCA}. Then, performing in-depth and user-friendly statistical analysis with massive datasets on a single-computer system becomes a problem of interest.

We thus argue that the interactive statistical analysis of massive datasets is important. Regardless of the size of a dataset (large or small), statistical analysis is often a journey of exploration with a lot of uncertainty. Without the results obtained from previous exploratory analyses, it is typically difficult to imagine what should be the next step. Should data be transformed in a certain way? Are there outliers need to be excluded? Is there a multicollinearity issue? What should be the best choice for the models (e.g., linear or nonlinear)? The answers to all these important questions depend on the results obtained from the early stage of analysis. One has to iteratively check the data and fix problems according to the feedback. We call this common practice an interactive statistical analysis. For applications with small data sizes, we seldom pay attention to this important issue because we take it for granted. Unfortunately, interactive statistical analysis becomes a luxury for massive datasets with limited computational resources. In this case, even for a task as simple as computing the sample mean, a long duration is required, making a user wait for results with often nothing else that can be done simultaneously. This process wastes time, resulting in a poor user experience.

The fundamental reason for such a long duration when performing these calculations on a personal computer is that a full data analysis must be performed for exploratory analysis. As we mentioned before, computing a statistic even as simple as the sample mean could be time-consuming if a massive dataset must be fully processed; this supposition immediately raises an interesting question: is this really necessary? The answer seems to be, ``No'', particularly for the early stage of the interactive statistical analysis. In this stage, users are most interested in making quick determinations about the data from different perspectives. However, in practical applications, these quick determinations can often be obtained with only a portion of the data. Therefore, a partial data-based convenient and quick solution is much more preferable than a time-consuming approach that requires all the data. Inspired by this, we designed a package to support the interactive analysis of massive datasets through subsampling. By subsampling, we mean that a subset of the entire dataset can be randomly sampled via simple random sampling with replacement. The subsample size is much smaller than the size of the original massive dataset. Thus, the sampled dataset can be easily loaded into the memory of a computer and then thoroughly studied. Such an approach will be cost-effective and can be performed for many replications (i.e., multiple subsamples) according to the user's specification. Thus, we have developed a toolbox with many interesting and useful statistical methods. This toolbox is an open-source Python package named {\it CluBear}.

The proposed package {\it CluBear} is to some extent similar to another popularly used R package called {\it bigmemory} \citep{bigmemoryweb,bigmemorypackage} but with clear differences. First, {\it bigmemory} accelerates the reading of data on disks by creating memory-mapped files (MMFs) in the memory of a computer and thus requires a continuous area of memory to retain MMFs. If the data is large, the corresponding MMFs may take a relatively large space in memory. In addition, {\it bigmemory} requires additional disk space that is often larger than the data size. When the memory or the disks do not satisfy these requirements, {\it bigmemory} must manage data block-by-block, which may slow down computations. In addition, if the original data change, the MMFs must be recreated. {\it CluBear} is based on subsampling conducted on the hard disk. Thus, {\it CluBear} has few requirements for either memory or disk space. Theoretically speaking, {\it CluBear} can manage files of arbitrary sizes if they can be placed on hard disks. Second, {\it bigmemory} uses a column-major format, which is convenient for column-based calculations. {\it CluBear} uses the row as the basic unit, which is more suitable for statistical analysis. Third, {\it bigmemory} itself lacks visualization and statistical tools, while {\it CluBear} has integrated graphs and statistical functions, which makes {\it CluBear} a convenient tool for interactive analysis.

The remainder of this article is organized as follows. Section~\ref{sec:methods} discusses the subsampling algorithm and the usage of the {\it CluBear} package. Section~\ref{sec:software} describes a case study that applies {\it CluBear} to an airline dataset. Section~\ref{sec:conclusion} concludes the paper.

\section{Interactive Statistical Analysis by Subsampling} \label{sec:methods}
\subsection{Why Interactive Statistical Analysis}

As mentioned before, statistical analysis in practice is often a journey that is full of uncertainty. This fact is particularly true for massive data with complex structures. As preliminaries for a formal analysis, a series of pre-processing steps must be considered. First, if the variables of interest are qualitative, their frequency tables must typically be created. Those classes with low frequencies or similar meanings are likely to be merged. In contrast, those classes with high frequencies might be subject to further partitioning. Second, if the variables of interest are quantitative, basic statistics must be computed. These basic statistics include but are not limited to the minimum, median, maximum, mean, standard deviation, and others. In addition, users might also want to graphically display the data by (for example) boxplots and histograms. All of these exploratory analyses must be conducted so that problems in the data can be detected as early as possible. To fix these problems, appropriate measures should be taken. This process often must be replicated many times because new problems are likely to be detected after the old problems are solved. Thus, this process calls for intensive interactions between the user and the statistical software. We call this a process of interactive statistical analysis (ISA), which is an important process for any complex statistical analysis. Existing literature pays little attention to ISA because the datasets used in traditional applications are often sufficiently small that all of the data can be easily read into the computer memory. Then, all interactive analyses can be quickly conducted without memory constraints. In such circumstances, ISA seems to be an easy or even trivial task, which has been always taken for granted. Unfortunately, this positive experience with ISA changes dramatically if massive datasets must be analyzed with limited memory.

Consider, for example, the airline dataset \citep{Airlinedata}, which describes U.S. flight information from October 1987 to April 2008. Airline data contains a total of approximately 123 million records, each of which contains flight date, flight time, flight ID, flight duration, airport information, unusual changes of the flight, and delay information. The entire dataset takes approximately 12 GB of the hard disk. For most typical computers with limited memory space, this dataset cannot be read into the memory as a whole. Thus, this dataset cannot be handled easily by traditional methods. To address the problem, we provide an example: computing the sample mean for the variable {\it ArrDelay}, which is a quantitative variable about the arrival delay in minutes. Because the entire airline dataset cannot be fully loaded into the memory, a Python program based on the divide-and-conquer strategy must be developed. The objective in this study is to partition the entire airline dataset into many small pieces (e.g., a total of 124 partitions with 1 million records per partition). Next, all partitions are processed sequentially, which takes approximately 6.6 minutes to obtain the final result about the sample mean on a computer with a 3.1-GHz Intel Core i7 CPU and 16 GB of memory, excluding the time cost due to program development and debugging. If the sample mean about the {\it ArrDelay} is the only statistic to compute, the time cost of approximately 6.6 minutes might be tolerable in practice. However, for a dataset as complex as that described above, many trials of the exploratory analysis must be performed before the final formal analysis is completed. As expected, the proposed exploratory analysis reveals many data problems. For example, the distribution of {\it ArrDelay} is particularly heavy-tailed. Thus, an appropriate data transformation must be applied before the formal analysis is performed. Many variables suffer from a large proportion of missing values. Thus, they must be dropped before the formal analysis. There are many other problems discovered via the exploratory analysis. Thus, we must add many tasks to the schedule before the formal analysis is performed. However, if each task (e.g., computing the sample mean) requires a sophisticated divide-and-conquer program to be developed, the entire analysis will take a long time to complete. As a result, the user experience about ISA could be poor.

One possible solution to this problem is distributed computing. If there is a powerful distributed computing system that consists of a large number of computers, the massive dataset can be fully loaded and processed with a markedly reduced time cost. The prominent examples in this regard are those distributed systems supported by Hadoop and Spark. Without any doubt, these systems provide good solutions but also suffer from at least two serious limitations. First, these systems are expensive. By definition, a powerful distributed system should contain many computers. Because so many computers are too expensive for most users, the maintenance of such a system, which is a complex aggregation of hardware and software, also requires considerable labor costs. Another possible solution is to rent a distributed system from a cloud platform (e.g., AWS, Azure, Aliyun). The financial cost can be reduced markedly but remains expensive. Second, this solution is complex. To make good use of those distributed systems, appropriate software support is required. For example, a distributed file system such as HDFS is typically necessary. Then, a distributed computing framework (e.g., Hadoop and Spark) must be incorporated. At least the resulting personal experience with this solution seems to suggest that learning and then being experienced with those pieces of software is difficult; this fact is particularly true for many users with limited coding experience. Thus, how to support ISA for most general users with only a personal computer and limited coding experience becomes an important problem to solve and thus motivates us to develop the {\it CluBear} package.

\subsection{Subsampling Theory} \label{subsec:subsampling}

To solve this problem, we propose a novel and useful software package, {\it CluBear}, which is a toolbox that was developed based on subsampling theory. Thus, the methods implemented by {\it CluBear} have solid theoretical support from the statistical literature about subsampling \citep{mahoney2011randomized,drineas2011faster,ma2014statistical,wang2018optimal,wang2019more,yu2020optimal,ma2020asymptotic}. For the most recent theoretical treatment about subsampling, we refer to \cite{Shuyuan2021} and \cite{10032639}. In this subsection, we consider a moment estimator as a concrete example to demonstrate the basic concepts of subsampling. The focus in this study is to (1) present the proposed method and explain why it is useful for ISA; and (2) discuss the supporting asymptotic theory.

We denote the entire dataset as $\mathcal{F}=\{i: 1 \leq i \leq N\}$, where $N$ is the sample size, and we assume that $\mathcal{F}$ is a massive dataset. Then, the size $N$ is sufficiently large that $\mathcal{F}$ cannot be loaded into computer memory as a whole. However, a subsample of $\mathcal{F}$ of a sufficiently small size can be processed easily. Specifically, we use $\mathcal{S}_{k} \subset \mathcal{F}$ to denote the subsample obtained in the $k$-th replication with $1< k < K$. We denote the size of $\mathcal{S}_{k}$ to be $n$, and each element of $\mathcal{S}_{k}$ is obtained by simple random sampling with replacement. Typically, $n$ cannot be too large; otherwise, the subsample cannot be loaded into the memory as a whole. However, $n$ cannot be too small either; otherwise, the subsample may lead to an estimator that has a poor statistical efficiency. Therefore, a carefully selected subsample size $n$ is critical. We let $\mu=E\left(X_{i}\right)$ denote the first-order moment of $X_{i}$. Assume that the second-order moment of $X_{i}$ is finite with $\sigma^2={\rm{var}}(X_{i})$ and assume that the parameter of interest is $\theta=\mathrm{g}(\mu)$ for a known function $\mathrm{g}(\cdot)$. We consider the first-order moment $\mu$ for simplicity. Essentially any order and number of moments can be comfortably handled by the proposed theory. Specifically, $\mathcal{S}_{k}$ can be easily sampled from $\mathcal{F}$. Multiple subsamples can also be obtained efficiently because most modern computers support multithreading jobs. Once $\mathcal{S}_{k}$ is obtained, the parameter of interest can be estimated as $\hat{\theta}_{k}=\mathrm{g}\left(\hat{\mu}_{k}\right)$. Then, multiple estimates are averaged to be $\bar{\theta}=K^{-1} \sum_{k=1}^{K} \hat{\theta}_{k}$.

This process immediately leads to an interesting question: what are the asymptotic properties of such an estimator compared with the global estimator $\hat{\theta}$? Thus, we also assume that the nonlinear transformation $\mathrm{g}(\cdot)$ is a sufficiently smooth function. Next, a standard Taylor expansion type technique can be readily applied as:
\begin{equation}
\bar{\theta} \approx \theta+\dot{\mathrm{g}}(\mu) \cdot K^{-1} \sum_{k=1}^{K}\left(\hat{\mu}_{k}-\mu\right)+0.5 \ddot{\mathrm{g}}(\mu) \cdot K^{-1} \sum_{k=1}^{K}\left(\hat{\mu}_{k}-\mu\right)^{2},\label{eq1}
\end{equation}
where $\dot{\mathrm{g}}(\mu)$ and $\ddot{\mathrm{g}}(\mu)$ are the first- and second-order derivatives of ${\mathrm{g}}(\mu)$ with respect to $\mu$, respectively. This equation (\ref{eq1}) suggests that the variance of $\bar{\theta}$ is primarily determined by $\dot{\mathrm{g}}(\mu) \cdot (\hat{\mu}-\mu)$. In particular, we have \citep{Shuyuan2021}
\begin{equation}\label{eq:var}	
	{\rm{var}}(\bar{\theta}) \approx \dot{\mathrm{g}}(\mu)^2\sigma^2\left(\frac{1}{N}+\frac{1}{nK}\right).
\end{equation}
This variance (\ref{eq:var}) is the same as that of the global estimator $\hat{\theta}=g(\hat{\mu})$ if $K$ is sufficiently large (i.e., $nK\gg N$).
However, the bias could be different due to the expectation of the second term in (\ref{eq1}), which is of $O(1/n)$ order. We recall that the global estimator is $O_p(1/\sqrt{N})$ consistent. Thus, if $n^2 \gg N$, the basis of $\bar{\theta}$ is negligible, and the subsample size should be much larger than $\sqrt{N}$, which seems to be a mild condition for most practical applications. To apply this concept, we consider the airline dataset again. The total sample size $N = 123,534,969$, and $\sqrt{N} \approx 11115$; then, $n=10^5$ appears sufficient.
{\red In addition, by the theory of \cite{10032639}, if the subsample size is sufficiently large in the sense that $n/\sqrt{N} \to \infty$ as $N \to \infty$, then the resulting estimator can be  statistically as effect as the full sample estimators as long as the number of subsamples is large enough.}
For a more detailed theoretical treatment of the subsampling size, we refer to \cite{Shuyuan2021} and \cite{10032639}.
{\red In practice, by the function ``seek'' in Python, we are able to find the addresses of the beginning and the end of the dataset on the hard drive. Therefore, we are able to make a quick estimation about the full sample size $N$. Thereafter, an appropriate value for the subsample size $n$ can be specified. }

\subsection{CluBear Architecture} \label{subsec:arch}

Subsampling provides a solid theoretical foundation for interactive analysis. However, how to implement it in a software package is difficult to determine. A carefully designed architecture is necessary. Specifically, for {\it CluBear}, we decompose the entire ISA process into three different steps: subsampling, data cleaning, and statistical analysis. Details about these three steps are discussed below.
The first step of {\it CluBear} is subsampling. This step plays a fundamental role in the ISA. The objective in this study is to build a bridge between the massive dataset on the hard disk and the computer memory. The purpose of this mechanism is to enable fast and convenient data access for subsampling. Intuitively, this process is similar to pumping water from a pond. The pond is similar to the massive dataset placed on the hard disk, and the water pump is a carefully designed function so that a subset of the data can be easily pumped (i.e., sampled) from the pond. Once the pump is set up, we can then perform efficient subsampling. However, the ``data'' generated by the pump is the raw data and thus cannot be used immediately for ``cooking''
(i.e., the next-stage formal analysis). Appropriate data cleaning is thus required, which may require a tank so that those raw data can be temporally held and then cleaned. Once the subsampled data are cleaned, we are ready for cooking (i.e., formal analysis). In this respect, {\it CluBear} provides a comprehensive toolbox for both descriptive analyses and graphical displays. The implemented descriptive statistics include the sample mean, its standard error, the sample standard deviation, the sample minimum, the sample median, the sample maximum, the sample skewness, the averaged noncentralized kurtosis, and the sample missing probability. The graphical tools support histograms, bar charts, boxplots and grouped boxplots.

\subsection{CluBear Design Overview} \label{subsec:design}

The whole package allows a total of three different data source types. They are respectively the CSV file type without a codebook (i.e., Type 1), the CSV file type with a codebook (i.e., Type 2), and the SQL database with detailed variable information (i.e., Type 3). By Type 1, the source data should be placed on the hard drive in a CSV format. No codebook is required. Consequently, the {\it CluBear} package cannot infer about the variable information (e.g. qualitative or quantitative) automatically. Instead, this has to be done manually. In contrast, the Type 2 requires a codebook for the source data. The detailed variable information, e.g., the list of quantitative variables, the list of variables that need to be dropped or the scale levels of qualitative variables, should be provided in the codebook. Consequently, the {\it CluBear} can infer about each variable automatically. Lastly, Type 3 requires the data source to be in a SQL format. So that determined variable information can be provided by the SQL database. Accordingly, the variable information also can be automatically inferred by the {\it CluBear} package.

{\red To make computationally efficient inferences for massive datasets, {\it CluBear} applies a sequential subsampling technique proposed in \cite{10032639}. It is an effective technique for speeding up the subsampling process on the hard drive. More specifically, for a massive dataset too large to be fully loaded into the computer memory as a whole, the subsampling operation has to be done by reading different lines of the data file on a hard drive in a fully random way. If the raw data is not randomly shuffled, a subsample with size $n$ has to be obtained by repeatedly moving the file pointer to $n$ random addresses on the hard drive, which correspond to $n$ different random data lines on the hard drive. Unfortunately, this is a very time-consuming process. In contrast, if the raw data is randomly shuffled already, the subsample can be obtained in a sequential way. In other words, the file pointer can start with a random initial position. Thereafter, the consecutive $n$ data lines can be read in a sequential manner. The time cost due to repeatedly placing the file pointer to random addresses on a hard drive is saved. This leads to a significant reduction in time cost. This speeding-up effect is the consequence of both random shuffling and sequential subsampling. By random shuffling only, we cannot speed up the subsampling process.
}

\begin{figure}
	\centering
	\includegraphics[width=1\linewidth]{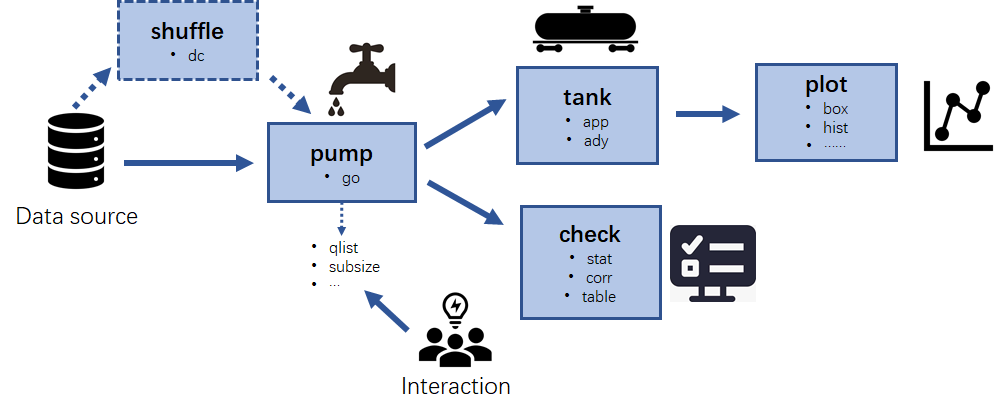}
	\caption{Design Flowchart of \it{CluBear}
	}\label{fig:flowchart}
	%	Calculate the marginal decision coefficient for each explanatory variable}
\end{figure}

The whole design structure of  the {\it CluBear} package is given in Figure~\ref{fig:flowchart}. It starts with the data source, which allows for three different types as mentioned before. Next, the data can be randomly shuffled by {\it dc}. By doing so, the sequential subsampling process can be conducted with a faster speed. However, this is not a must request. Even if the data are not randomly shuffled, the {\it pump} can be directly used to do subsampling. For the data source Type 2 and 3, the variable information can be automatically inferred. However, if the data source type is Type 1, then the variable information has to be inferred manually. This can be done by {\it check}. Once the data variable information is clear, it can be further cleaned by {\it tank}. If {\it pump} is considered as a ``faucet'', then {\it tank} is a ``reservoir''. After accumulating data in the ``reservoir'', operations such as data changes can be performed. Next, {\it plot} is used to draw pictures, such as using {\it box} and {\it hist} to draw box plots and histogram, respectively. Here, the forms of graphs can be added freely and are extensible.

\section{Airline data management} \label{sec:software}

\subsection{Airline Data and CluBear}

We next use an airline dataset to demonstrate the usefulness of {\it CluBear} for ISA. The airline dataset can be publicly downloaded from the official website of the ASA at https://community.amstat.org/jointscsg-section/dataexpo/dataexpo2009. The data was collected from all U.S. flight information from October 1987 to April 2008. The total sample size is approximately 123 million records, occupying approximately 12 GB of space on the hard disk. Each sample corresponds to one flight record within the United States. The following variables are included in the airline dataset:

\begin{itemize}
\item Date information, including 5 variables: (1) {\it Year}, (2) {\it Month}, (3) {\it DayofMonth}, and (4) {\it DayOfWeek}, where 1 means Monday and 7 means Sunday.
\item The local time information, including 4 variables: (1) {\it DepTime} (Actual departure time), (2) {\it CRSDeptTime} (Planned departure time), (3) {\it ArrTime} (Actual arrival time), and (4) {\it CRSArrTime} (Planned arrival time).
\item The flight operation information, including (1) {\it UniqueCarrier} (The airline's unique identification number), (2) {\it FlightNum} (The flight number), and (3) {\it TailNum} (the aircraft tail number), which is a unique identification number for the aircraft.
\item Time cost information (in minutes), including (1) {\it ActualElapsedTime} (Actual consumption time), (2) {\it ArrTime} (the real arrival time), (3) {\it DepTime} (the real departure time), (4) {\it CRSElapsedTime} (The
%Quality Control Editor: Please ensure that the intended meaning has been maintained in the following edit.
planned travel time
), (5) {\it CRSArrTime} (the planned arrival time), (6) {\it CRSDepTime} (the planned departure time), and (7) {\it AirTime}
(the real flight time of the aircraft in the air), (8) {\it Taxiout} (the take-off taxi time), and (9) {\it Taxiin} (the landing taxi time).
\item The airport information includes (1) {\it Origin} (The origin airport), (2) {\it Dest} (The destination airport), and (3) {\it Distance} (The flight distance between the two airports in miles). Very detailed supplementary information is also provided, including {\it country}, {\it city}, {\it long} (longitude), and {\it lat} (latitude) coordinates for each airport.
\item Unusual schedule changes, including (1) {\it Cancelled} (Whether the flight is cancelled), (2) {\it CancellationCode} (The reason for the cancellation), and (3) {\it Diverted} (Whether the flight is diverted).
\item Delay record in minutes, including (1) {\it ArrDelay} (Arrival Delay), (2) {\it DepDelay} (Departure Delay), (3) {\it CarrierDelay} (Airline Delay), (4) {\it SecurityDelay}, (5) {\it WeatherDelay}, (6) {\it NASDelay}, and (7) {\it LateAircraftDelay}.
\end{itemize}

Because there is so much data, they cannot be read into the memory as a whole. We thus consider analyzing the dataset using {\it CluBear}. A compatible computer environment that supports Python3 is required. In a shell window, {\it CluBear} can be easily installed using the following command:
\begin{lstlisting}
# pip install CluBear
\end{lstlisting}
\subsection{The Pump and Tank}

We next show how to build a bridge between the airline data and the memory using a pump. With the help of {\it CluBear}, this can be easily implemented as follows:
\begin{lstlisting}
>>> import clubear as cb
>>> mypathfile='airline.csv.shuffle'
>>> pm=cb.pump(type=1, pathfile=mypathfile)
>>> pm.seq=True
>>> pm.subsize=10**5
>>> pm.go()

    ActualElapsedTime   AirTime   ArrDelay   ArrTime    ...
0                96.0        NA        5.0    1410.0    ...
1                65.0        46       -4.0     912.0    ...	
2                86.0        NA       22.0    1850.0    ...	
3               202.0       185      -11.0    1745.0    ...
4               232.0       218       -2.0    1248.0    ...	
...               ...       ...        ...       ...    ...	
99995           106.0        84       -4.0     857.0    ...	
99996            75.0        61      -10.0     930.0    ...	
99997            77.0        66      -14.0    1233.0    ...	
99998           148.0        NA       18.0    1140.0    ...
99999            70.0        53        0.0    1610.0    ...	
100000 rows * 30 columns
\end{lstlisting}
In this code, a randomly shuffled airline dataset ``airline.csv.shuffle'' has been placed under the default directory. As a result, a pump can be easily established by the {\it cb.pump} command. Next, by {\it pm.size=10**5}, we set the subsample size to be $10^5$. Then, a subsample of this size can be easily obtained by the {\it pm.go} function. Note that ``airline.csv.shuffle'' here is a CSV file type without a codebook (i.e., Type 1), which can be handled by the third line of the above code. For the simplicity of replication, we prepared a mini-version of the airline data ``2008.csv'', which includes about 701w rows of records for airline data in 2008. The randomly shuffled file is ``2008.shuffle.csv''. If a codebook can be provided for the mini-version data, then we can establish Type 2 pump as follows. Here {\it qlist}, {\it drop}, {\it scale\_level} specify the list of quantitative variables, the list of variables that need to be dropped, and the scale levels of qualitative variables, respectively.
\begin{lstlisting}
>>> mycodebook={
	'qlist': ['ActualElapsedTime', 'ArrDelay', 'ArrTime', 'CRSArrTime',
	          'CRSDepTime', 'CRSElapsedTime', 'Cancelled', 'DayOfWeek',
	          'DayofMonth', 'DepDelay', 'DepTime', 'Distance', 'Month',
	          'FlightNum', 'Year', 'Diverted', '_INTERCEPT_'],
	'drop': ['TailNum', 'Origin', 'Dest'],
	'scale_level': {'UniqueCarrier': ['EV', 'OO', '9E']}}
>>> mypathfile='2008.csv.shuffle'
>>> pm=cb.pump(type=2, pathfile=mypathfile, codebook=mycodebook)
\end{lstlisting}
		
If the mini-version data were stored in a SQL database (e.g., stored in a table named ``flight\_2008'' within a database named ``test\_db''), then the Type 3 data loading pump can be enforced as follows:
\begin{lstlisting}
>>> mysql_info={
	'host':'localhost',
	'user':'root',
	'passwd':'root',
	'port':3306,
	'db':'test_db',
	'tablename':'flight_2008'
    }
>>> pm=cb.pump(type=3, sql_info=mysql_info)
\end{lstlisting}
	
It should also be noted that our approach allows both shuffled and non-shuffled dataset. Take Type 1 mini-version data for example. If the data are not shuffled, our approach remains applicable without any problem. A program illustration for the non-shuffled mini-version data as follows, where the value of {\it seq} is set to be {\it False}.	

\begin{lstlisting}
>>> mypathfile='2008.csv'
>>> pm=cb.pump(type=1, pathfile=mypathfile)
>>> pm.seq=False
\end{lstlisting}

By the theory of \cite{Shuyuan2021}, we obtain the asymptotic variance formula (\ref{eq:var}). Therefore, we know that the variance of the subsampled estimator is mainly determinated by $n$ and $K$ in the form of (\ref{eq:var}). Consequently, for a practical implementation, the subsample size  $n$ should be taken as large as possible, as large as the computer memory capacity allows. With a given $n$, the total number of subsamples $K$ should also be taken as large as possible so that the estimated SE value falls below pre-specified level. Fortunately, the SE of a resulting estimator can be automatically provided by {\it CluBear}. Note that the output SE is the real SE value multiplied by 100 for concise presentation. Take the {\it Distance} indicator of the mini-version dataset as an example. With $n=10^5$ and $K = 5$, the SE is about 0.80 miles; see the detailed output in the program below.

\begin{lstlisting}
>>> pm.subsize=10**5
>>> cb.check(pm).stats('Distance', niter=5)

Time elapsed: 18.9 seconds with subsample sizes 100000 .
Task accomplished:  100.0 % for a total of  5 random replications.
		
              Mu     SE     Std   Min    Med     Max  Skew  Kurt   mp
Distance  731.62  79.90  564.99  21.0  585.0  4962.0  1.63  6.27  0.0
\end{lstlisting}

{\red The option $niter$ is used to specify the number of subsamples $K$.} If we wish to have an SE smaller than 0.40 miles, then we can consider increasing $K$ to 20. Accordingly, the SE falls below 0.40 miles; see the detailed output in the program as follows. %By doing so, the smallest $K$ value can be determined.

\begin{lstlisting}
>>> cb.check(pm).stats('Distance', niter=20)

Time elapsed: 77.4 seconds with subsample sizes 100000 .
Task accomplished:  100.0 % for a total of  20 random replications.

              Mu     SE     Std   Min    Med     Max  Skew  Kurt   mp
Distance  731.35  39.86  563.76  21.0  585.0  4962.0  1.62  6.23  0.0
\end{lstlisting}

{\red
Regarding the selection of $K$, we have two different strategies.
First, we can monitor the reported SE from the console of the subsampling process. When the reported SE becomes smaller than a pre-specified precision level, we can then stop the estimation process. By doing so, the number of subsamples $K$ can be automatically determined.
The second strategy is to manually specify a value for $K$ so that $nK \gg N$, if we are given sufficient amount of computation time. Then, by the theory of \cite{10032639}, the resulting estimatior should be statistically as efficient as the full sample estimator.
}

{\red
It is noteworthy that subsampling-based estimation methods, such as those implemented by {\it CluBear}, are not computationally cheaper than those total-sample-based methods.
The reason is simple. By subsampling, additional noises are artificially created, which cannot be beneficial in terms of computation cost. Therefore, computation costs can never be saved by subsampling. Then, a natural question is: Why do we still prefer subsampling? We prefer subsampling for two reasons. The first reason is that subsampling methods improve the computational ``feasibility'', when researchers are given very limited computation resources. In this case, the whole sample analysis is extremely difficult in practice or even infeasible, even though its overall computation cost is less than that of the subsampling-based method. Therefore, the subsampling technique improves the computational feasibility of massive data analysis. The second reason is that for many real applications, the practical demand for the estimation accuracy is actually very limited. In this case, the estimation accuracy provided by the whole sample analysis is often too luxurious and time-consuming to be practically necessary. In fact, obtaining a slightly less accurate estimate but with a significantly reduced time cost is often practically more appealing. Therefore, the subsampling technique improves the computational flexibility for massive data analysis.}

{\red
To summarize, the computation cost of the subsampling method is not cheap at all, as compared with the estimator based on the total sample, if the same level of estimation accuracy is needed. However, the subsampling method greatly improves the computational feasibility and flexibility. We take the calculations of the mean and the median of {\it Distance} for example. The estimation based on the whole dataset (``2008.csv.shuffle'') requires 39.9 seconds. The sample mean and the sample median are 726.39 miles and 581 miles, respectively. Then, we apply the sequential subsampling method to estimate the mean and the median. Consider for example $n=10^5$ and $K = 10$, the subsampling-based estimates of the mean and the median are given by 726.34 miles and 581 miles, respectively. They are very close to the total-sample-based results. The total time cost is 20.2 seconds, which is much cheaper than that of the total-sample-based method.
\begin{lstlisting}
>>> mypathfile='2008.csv.shuffle'
>>> pm=cb.pump(type=1, pathfile=mypathfile)
>>> pm.subsize=10**5
>>> pm.seq=True
>>> cb.check(pm).stats('Distance', niter=10)

Time elapsed: 20.2 seconds with subsample sizes 100000 .
Task accomplished:  100.0 % for a total of  10 random replications.

              Mu     SE     Std   Min    Med     Max  Skew  Kurt   mp
Distance  726.34  56.26  562.58  21.0  581.0  4962.0  1.64  6.28  0.0
\end{lstlisting}
A more detailed comparison between the sequential subsampling method and the total-sample-based method is given in Figure \ref{fig:compare}, which displays the subsampling-based estimates (blue solid lines) with different time budgets and the total-sample-based estimates (red dashed lines). Consider a case where the time budget is only 5 seconds. In regard to the mean of {\it Distance}, the absolute difference between the subsampling-based estimate (726.57 miles) and the total-sample-based estimate (726.39 miles) is only 0.18 miles. For the median of {\it Distance}, the absolute difference is also close to 0. In this case, compared to the total-sample-based method, the subsampling-based method achieves similar results but with a significantly reduced time cost.
\begin{figure}
	\centering
	\includegraphics[width=0.8\linewidth]{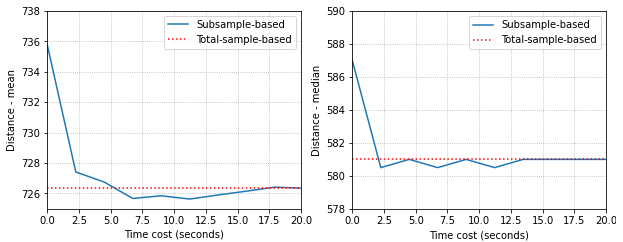}
	\caption{\label{fig:compare}
		Comparisons between the subsampling-based estimation and the total-sample-based estimation in terms of the mean (left) and the median (right) of {\it Distance}}
\end{figure}
}
	
\subsection{Descriptive Statistics} \label{subsec:software-tank}
Once pump {\it pm} is correctly set up, descriptive statistics can be easily computed. To achieve this, we must first create a {\it cb.check} object on {\it pm}. The {\it check} class is specifically designed for inspecting subsamples pumped by {\it pm}. Then, various statistics, frequency tables and correlation matrixes can be produced. {\red By calling the {\it stats} function, a group of the most frequently used descriptive statistics, including sample variance, skewness and kurtorsis, can be computed as:
\begin{lstlisting}
>>> ck=cb.check(pm).stats()

Time elapsed: 42.5 seconds with subsample sizes 69734.
Task accomplished:  100.0 % for a total of  10 random replications.

            Mu     SE    Std     Min    Med    Max  Skew   Kurt    mp
ActualE... 128.0   7.1   70.4   14.0  110.0  772.0   1.5    5.7   2.2
AirTime    104.7   6.8   67.7    0.0   87.0  753.0   1.5    5.9   2.2
ArrDela... 8.1     3.9   38.7  -81.0  -2.0  1951.0   5.5   77.8   2.2
ArrTime    1484.1  51.0  504.9   1.0 1514.5 2400.0  -0.3    2.6   2.1
CRSArrT... 1498.0  48.2  482.0   0.0 1520.0 2400.0  -0.2    2.4   0.0
CRSDepT... 1327.8  46.4  463.9   1.0 1320.0 2359.0   0.1    2.0   0.0
CRSElap... 129.6   7.0   69.7  -141.0 112.0 1435.0   1.5    6.0   0.0
Cancell... 0.0     0.0   0.1     0.0    0.0    1.0   7.0   50.0   0.0
Carrier... 15.8    8.7   40.6    0.0    0.0 1951.0   8.4  149.8  78.3
DayOfWe... 3.9     0.2   2.0     1.0    4.0    7.0   0.1    1.8   0.0
DayofMo... 15.8    0.9   8.8     1.0   16.0   31.0   0.0    1.8   0.0
DepDela... 10.0    3.6   35.8   -64.0  -1.0 1952.0   6.6  104.2   1.9
DepTime    1335.5  48.2  477.8   1.0 1327.0 2400.0   0.1    2.0   2.0
Dest       nan     nan   nan     nan    nan    nan   nan    nan 100.0
Distanc... 732.3   56.5  564.5   11.0 586.5 4962.0   1.6    6.3   0.0
Diverte... 0.0     0.0   0.1     0.0    0.0    1.0  20.2  408.3   0.0
FlightN... 2222.0  196.1 1961.4  1.0 1569.0 9743.0   0.9    2.8   0.0
LateAir... 20.8    8.4   39.4    0.0    0.0  850.0   3.6   23.9  78.3
Month      6.4     0.3   3.4     1.0    6.0   12.0   0.0    1.8   0.0
NASDela... 17.2    6.9   32.1    0.0    6.0 1337.0   5.2   64.7  78.3
Origin     nan     nan   nan     nan    nan    nan   nan    nan 100.0
Securit... 0.1     0.4   1.8     0.0    0.0  357.0  43.0 2618.2  78.3
TailNum    nan     nan   nan     nan    nan    nan   nan    nan 100.0
TaxiIn     6.9     0.5   5.0     0.0    6.0  184.0   4.6   55.4   2.1
TaxiOut    16.5    1.1   11.3    0.0   14.0  393.0   4.6   50.3   1.9
UniqueC... nan     nan   nan     nan    nan    nan   nan    nan 100.0
Weather... 3.0     4.2   19.6    0.0    0.0 1225.0  13.5  318.4  78.3
Year       2008.0  0.0   0.0  2008.0 2008.0 2008.0   0.0    0.0   0.0
_INTERC... 1.0    0.0    0.0     1.0    1.0    1.0   0.0    0.0   0.0

* Mu: the averaged subsample mean;
* SE: the standard error of Mu in %;
* Std: the averaged subsample standard deviation;
* Min: the minimum of the subsample minimum;
* Med: the median of the subsample median;
* Max: the maximum of the subsample maximum;
* Skew: the averaged sample skewness;
* Kurt: the averaged non-centralized kurtosis;
* mp: the averaged subsample missing probability in percentage (0%~100%).
\end{lstlisting} }
Based on the {\it mp} (missing probability) values in the last column, we find that those variables can be categorized into three groups. The first group contains variables with {\it mp} values near 0. A variable with such a characteristic is likely a quantitative value, although it may have a small percentage of missing values. For example, the {\it mp} value of {\it ArrDelay} is only 2.2\%. The second group contains variables with large {\it mp} values. As shown in the results, these data make more than 90\%, or even 100\% of the data. If this happens, the corresponding variables are more likely to be qualitative variables because their values in the original data are the most characteristic and thus cannot be easily turned into actual quantitative data. This issue leads to large {\it mp} values; for example, the {\it mp} of {\it CancellationCode} is 100.0\%. The remaining variables form the third group. In this case, a variable in this group is likely to be a quantitative variable but subject to a large percentage of missing values.

{\it CluBear} treats all variables as qualitative by default. The above statistical result suggests that a number of variables are supposed to be quantitative. To guide {\it CluBear} to distinguish between quantitative and qualitative variables, we must provide {\it CluBear} with a list containing quantitative variables and pass this information to the {\it qlist} parameter. Then, {\it CluBear} will only recognize those variables in {\it qlist} as quantitative and other variables as qualitative. Specifically, we use the following code to extract variables whose {\it mp} values are less than 5\%. Those variables are then treated as quantitative:
\begin{lstlisting}
>>> ck[ck.mp<5].index

Index(['ActualElapsedTime', 'AirTime', 'ArrDelay', 'ArrTime',
        'CRSArrTime','CRSDepTime', 'CRSElapsedTime', 'Cancelled',
       'DayOfWeek', 'DayofMonth','DepDelay', 'DepTime', 'Distance',
        'Diverted', 'FlightNum','Month','TaxiIn', 'TaxiOut',
        'Year', '_INTERCEPT_'], dtype='object')
\end{lstlisting}
Using command {\it ck[ck.mp$<$5].index}, we collect all those variables with miss probability (i.e., {\it mp}) values less than 5\%. Those variables are most likely to be quantitative. We then pass this list to {\it pm} by the {\it qlist} parameter:
\begin{lstlisting}
>>> pm.qlist=['ActualElapsedTime', 'AirTime', 'ArrDelay', 'ArrTime',
       'CRSArrTime', 'CRSDepTime', 'CRSElapsedTime', 'Cancelled',
        'DayOfWeek','DayofMonth', 'DepDelay', 'DepTime',
      'Distance', 'Diverted', 'FlightNum', 'Month', 'TaxiIn',
       'TaxiOut', 'Year', '_INTERCEPT_']
\end{lstlisting}

We next consider how to conduct a descriptive analysis for qualitative variables. Frequency tables can thus be produced. However, to generate a frequency table for a categorical variable, we must know how many levels are involved. Different numbers of levels are likely to be included by different subsamples. Thus, a large number of subsamples might be required to discover all levels; this can be done easily by the {\it table} function as:

\begin{lstlisting}
>>> tb=cb.check(pm).table(niter=20)

Time elapsed: 272.1 seconds.
Task Completed:  100.0 % for a total of 20 iterations for totally
2000000 counts.

[ 00 ] No. of levels detected for * CancellationCode  * is: 6
[ 01 ] No. of levels detected for * CarrierDelay      * is: 683
[ 02 ] No. of levels detected for * Dest              * is: 304
[ 03 ] No. of levels detected for * LateAircraftDelay * is: 476
[ 04 ] No. of levels detected for * NASDelay          * is: 450
[ 05 ] No. of levels detected for * Origin            * is: 303
[ 06 ] No. of levels detected for * SecurityDelay     * is: 101
[ 07 ] No. of levels detected for * TailNum           * is: 5360
[ 08 ] No. of levels detected for * UniqueCarrier     * is: 21
[ 09 ] No. of levels detected for * WeatherDelay      * is: 446
\end{lstlisting}
In this code, the function {\it table} is used to discover the number of levels for each variable. The option {\it niter=20} specifies the number of subsamples to be generated. We find that different categorical variables might have different numbers of levels. For example, {\it CancellationCode} has only 6 levels. In contrast, {\it Dest} and {\it Origin} have more than 300 levels discovered. For illustration purposes, we can drop these variables with too many levels by passing this information to the {\it drop} parameter of {\it pm} as follows:
\begin{lstlisting}
>>> pm.drop=['TailNum', 'Origin', 'Dest']
\end{lstlisting}
After those three variables (i.e., {\it TailNum}, {\it Origin}, {\it Dest}) with too many levels dropped, we can then regenerate the frequency table for {\it AirTime} as:
\begin{lstlisting}
>>> tb=cb.check(pm).table('AirTime',tv=True)

Time elapsed: 44.6 seconds.
Task Completed:  100.0 % for a total of 10 iterations for totally
1000000 counts.

[ 0 ] No. of levels detected for * AirTime * is: 607
nan = 2.21 %|45 = 0.98 %|51 = 0.97 %|44 = 0.97 %|46 = 0.97 %|
55 = 0.96 %|49 = 0.96 %|53 = 0.96 %|50 = 0.95 %|56 = 0.95 %|
48 = 0.95 %|52 = 0.95 %|54 = 0.95 %|47 = 0.94 %|59 = 0.93 %|
43 = 0.92 %|60 = 0.92 %|58 = 0.91 %|57 = 0.90 %|61 = 0.90 %|
63 = 0.90 %|62 = 0.89 %|64 = 0.89 %|65 = 0.88 %|42 = 0.88 %|
66 = 0.88 %|67 = 0.86 %|41 = 0.84 %|68 = 0.82 %|69 = 0.82 %|
40 = 0.79 %|70 = 0.78 %|71 = 0.78 %|72 = 0.77 %|73 = 0.74 %|
85 = 0.72 %|75 = 0.72 %|80 = 0.72 %|83 = 0.72 %|74 = 0.71 %|
39 = 0.71 %|84 = 0.71 %|86 = 0.71 %|78 = 0.71 %|76 = 0.71 %|
82 = 0.70 %|79 = 0.70 %|81 = 0.70 %|77 = 0.69 %|87 = 0.69 %|
88 = 0.68 %|89 = 0.67 %|90 = 0.67 %|91 = 0.66 %|93 = 0.66 %|
38 = 0.64 %|92 = 0.64 %|37 = 0.63 %|94 = 0.62 %|95 = 0.61 %|
96 = 0.59 %|97 = 0.59 %|36 = 0.59 %|35 = 0.59 %|100 = 0.57 %|
99 = 0.57 %|98 = 0.57 %|101 = 0.55 %|34 = 0.54 %|103 = 0.54 %|
105 = 0.53 %|102 = 0.53 %|33 = 0.52 %|107 = 0.51 %|106 = 0.51 %|
104 = 0.51 %|32 = 0.51 %|109 = 0.50 %|108 = 0.50 %|111 = 0.50 %|
110 = 0.49 %|115 = 0.49 %|112 = 0.49 %|114 = 0.48 %|31 = 0.48 %|
120 = 0.48 %|30 = 0.48 %|113 = 0.47 %|121 = 0.47 %|122 = 0.47 %|
125 = 0.47 %|118 = 0.47 %|117 = 0.47 %|116 = 0.47 %|124 = 0.47 %|
119 = 0.47 %|123 = 0.46 %|126 = 0.46 %|128 = 0.46 %|129 = 0.46 %|
*** Note: only top 100 levels displayed...
\end{lstlisting}
In this code, the option {\it tv} is an abbreviation for ``table view''. By setting {\it tv=True}, the {\it table} function will report the relative frequency of each level for the target variable. Because the variable {\it AirTime} is a quantitative variable in nature, the number of its levels is too large. In this situation, {\it CluBear} is designed to show only the top 100 levels with the highest percentages. {\red The result intuitively indicates that the missing values of AirTime are all marked as {\it nan}.}

\subsection{Statistical Graphics} \label{subsec:software-visual}
The {\it CluBear} package provides various functions for statistical graphics that are all included in the class {\it plot}. The supported statistical graphics are histograms ({\it hist}), bar charts about the mean ({\it mu}), bar charts about the standard deviation ({\it std}), bar charts for the frequency of different classes ({\it size}), heatmaps for the correlation matrix ({\it corr}), boxplots ({\it box}) and grouped boxplots ({\it gbox}). Details of these plots are discussed below.

\begin{figure}
	\centering
	\includegraphics[width=0.7\linewidth]{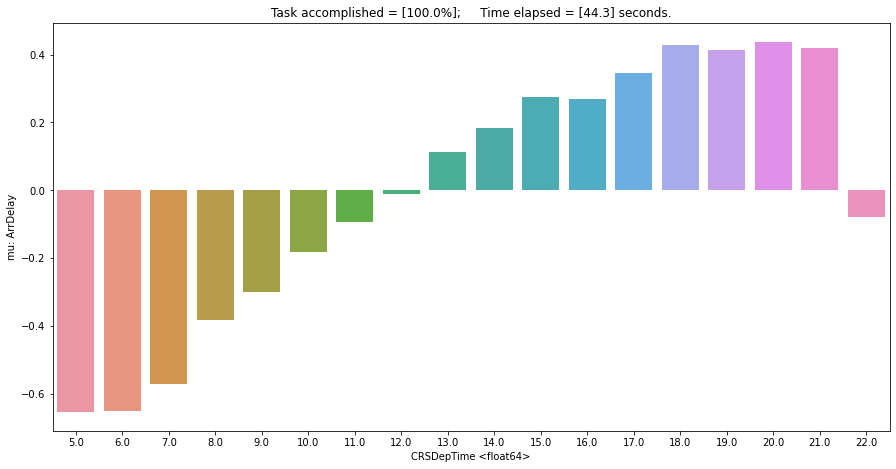}
	\caption{\label{fig:bar}
		The barplot of the means of \it{ArrDelay} according to the discretized \it{CRSDepTime}}
\end{figure}

For illustration purposes, we consider cutting the data into different groups according to the value of {\it CRSDepTime}, which can be done easily by the function {\it app}. Specifically, we generate a tank {\it tk} with a pump {\it pm} as the input. Then, a user-defined {\it lambda} function is applied to the {\it tk} object so that the originally continuous variable {\it CRSDepTime} can be transformed into a discrete variable. Next, two {\it app} operations are used to restrict the range of {\it CRSDepTime} to within the range of $[5,22]$. Then, we apply a logarithmic transformation to {\it ArrDelay} for illustration purposes. Because {\it ArrDelay} can be negative (early arrival), one particular type of transformation is used: $sign(x)*log(1+|x|)$. This transformation retains both the scale and the sign information of the original data. Last, we compute the means of {\it ArrDelay} according to the discretized {\it CRSDepTime}, where the means corresponding to different levels of {\it CRSDepTime} are visualized in bar plots by the {\it mu} function. Figure \ref{fig:bar} shows the result. A clear pattern can be detected in Figure \ref{fig:bar}. Early flights (before 10:00 am) are less likely to be delayed. In contrast, flights at approximately 19:00 tend to have higher delay risks.
\begin{lstlisting}
>>> import numpy as np
>>> tk=cb.tank(pm)
>>> tk.app(lambda x: np.floor(x/100),'CRSDepTime')
>>> tk.app(lambda x: max(x,5),'CRSDepTime')
>>> tk.app(lambda x: min(x,22),'CRSDepTime')
>>> tk.app(lambda x: np.sign(x)*np.log(1+np.abs(x)),'ArrDelay')
>>> pt=cb.plot(tk).mu('ArrDelay','CRSDepTime')
\end{lstlisting}
Next, we consider how to make grouped boxplots for {\it ArrDelay} according to {\it Month}. We wish to cut the original sample into different groups according to the values of {\it Month}. Next, boxplots are provided for {\it ArrDelay} for every {\it Month}. Figure \ref{fig:box} displays the resulting boxplots. This can be done easily by the {\it box} function as:
\begin{lstlisting}
>>> pt=cb.plot(tk).box(x="Month",y="ArrDelay")
	\end{lstlisting}
\begin{figure}
	\centering
	\includegraphics[width=0.7\linewidth]{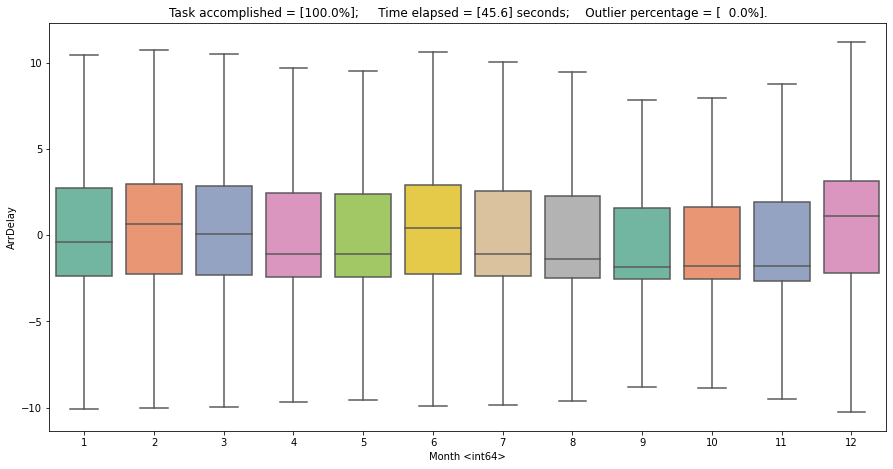}
	\caption{\label{fig:box}
		The boxplots of \it{ArrDelay} for every \it{Month}}
\end{figure}

We find that the variability of {\it ArrDelay} (Arrival delay) does not vary much across different {\it Month}. In contrast, clear median differences can be detected. We next consider how to produce histograms and take the variable {\it Distance} as an example. This variable is heavy-tailed and random with only positive values. We thus first apply a {\it np.log} transformation to it by the function {\it app}. Next, we use {\it hist} to make an interactive histogram as follows:
\begin{lstlisting}
>>> tk.app(np.log, 'Distance')
>>> pt=cb.plot(tk).hist('Distance')
	\end{lstlisting}
\begin{figure}
	\centering
	\includegraphics[width=0.7\linewidth]{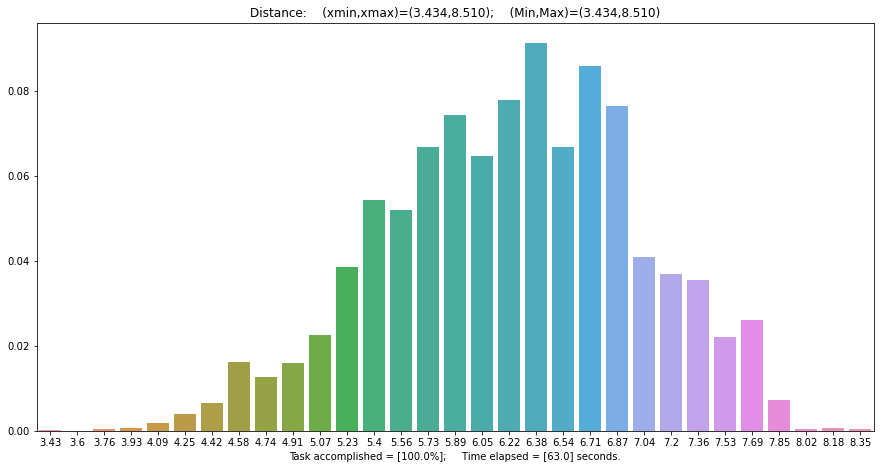}
	\caption{\label{fig:hist}
		The histogram of \it{Distance}}
\end{figure}

According to the histogram shown in Figure \ref{fig:hist}, we find that the distribution of the log-transformed {\it Distance}
%Quality Control Editor: Please note that some text appears to be missing here. Please consider adding any missing information.
is in a good shape. Most flight distances are concentrated between 5.73 and 6.87 in log values (307.97 and 962.95 miles, respectively). The proportions of short-distance and long-distance flights are both small.

\begin{figure}
	\centering
	\includegraphics[width=0.8\linewidth]{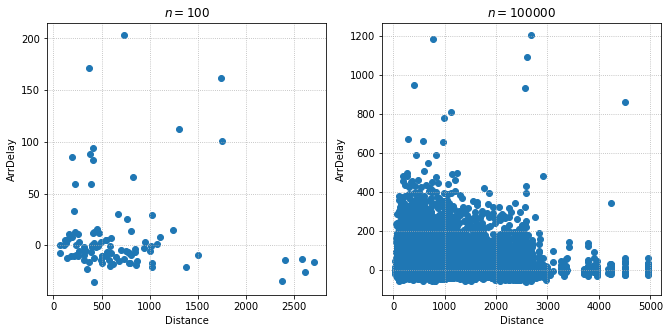}
	\caption{\label{fig:scatter}
		The scatter plot for {\it Distance} and {\it ArrDelay} with $n=100$ (left) and that of $n=100000$ (right)}
\end{figure}

We next consider how to graphically show the correlation relationship between two quantitative random variables. With traditional practice, various scatter plots are typically used.
%However, we find this method less effective for massive data analysis because the sample size in this study is too large. The resulting scatter plot may be less informative because the graph could be messy.
{\red However, one subsampling-based scatter might not be sufficient to provide a complete picture about the data pattern, if the subsample size $n$ is too small, as is shown in the left panel of Figure \ref{fig:scatter}. On the other side, with a fixed figure size, an extremely large subsample size will lead to many overlapped data points, as is shown in the right panel of Figure \ref{fig:scatter}. This makes the resulting figure even less informative. Therefore, we advocate to use multiple subsample plots with an appropriate subsample size for graphical inspection. By appropriate subsample size, we can assure each data point can be well displayed in the figure. By repeatedly inspecting many subsampling-based plots, we can assure that the generated subsample plots collectively to be sufficiently representative for the whole sample.}

\begin{figure}
	\centering
	\includegraphics[width=0.7\linewidth]{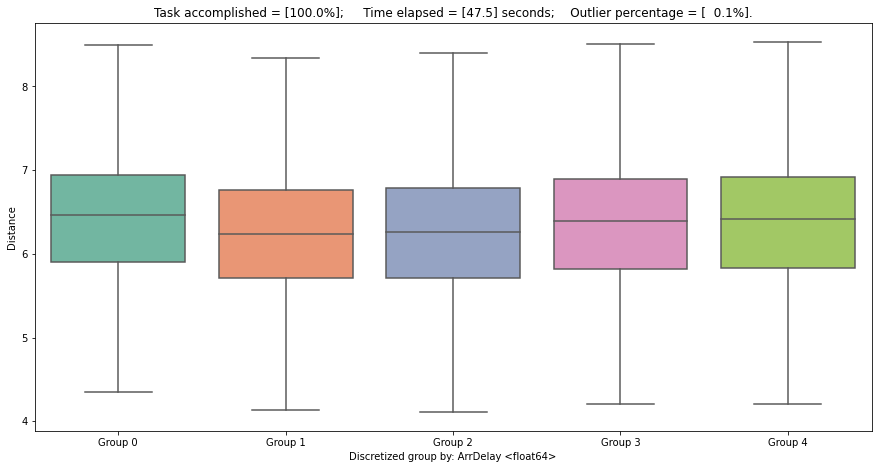}
	\caption{\label{fig:mrs}
		The grouped boxplot for \it{Distance} and the discretized \it{ArrDelay}}
\end{figure}

As an alternative solution, we treat one quantitative variable as a covariate $x$ and the other as a response variable $y$. Then, we discretize $x$ into equally sized groups. Then, a grouped boxplot can be generated for $y$. For convenience, we refer to this as the grouped boxplot method. We find this method to be much more reliable and effective than the conventional scatter plot. This method can also be conducted readily by the {\it gplot} function as follows. Figure \ref{fig:mrs} illustrates the result. Based on the output figure, no clear correlation between flight distance and delay can be visually detected.
\begin{lstlisting}
>>> pt=cb.plot(tk).gbox(y='Distance',x='ArrDelay')
	\end{lstlisting}

{\red
\subsection{Statistical Models}
The {\it CluBear} package supports more complex analyses such as linear regression and logistic regression. The modeling functions are provided in the class {\it model}. A linear regression analysis can be readily implemented by the {\it ols} function. For example, we construct a linear regression model, where the dependent variable is {\it ArrDelay} (Arrival delay) and the independent variables are {\it DepDelay} (Departure delay) and {\it Distance} (The flight distance between the two airports in miles).
\begin{lstlisting}
>>> mycodebook={'qlist': ['ArrDelay', 'DepDelay', 'Distance', '_INTERCEPT_']}
>>> mypathfile='2008.csv.shuffle'
>>> pm=cb.pump(type=2, pathfile=mypathfile, codebook=mycodebook)
>>> pm.subsize=10**5
>>> pm.seq=True
>>> md=cb.model(pm).ols('ArrDelay', niter=20, tv=True)

Time elapsed: 31.9 seconds with averaged subsample sizes 97789.1 and R.Squared =  86.8 %
Task accomplished:  100.0 % for a total of  20 random replications.

             Estimate  StandErr    tStat  pValue
DepDelay        1.019     0.028 3585.779   0.000
Distance       -0.001     0.002  -69.370   0.000
_INTERCEPT_    -1.047     1.652  -63.395   0.000

* Estimate: the ordinary least squares estimate for standardized x variables.
* StandErr: the standard error in %; t.Stat: the t-statistics = Estimate/Stand.Err.
* pValue: the p-value computed by normal approximation in %.
\end{lstlisting}
In this code, we use a codebook to specify the list of required quantitative variables, including both dependent and independent variables. To conduct the ordinary least square estimation for our linear regression model, we adopt the {\it ols} function and specify the name of the dependent variable ``ArrDelay''. By setting the option {\it tv=True}, the regression result will be displayed in a table. In this case, the $p$-values for the regression coefficients of {\it DepDelay} and {\it Distance} are highly close to 0, which means that those coefficients are statistically significant. In addition, the value of R.Squared is relatively high (86.8\%), which means that the variation of  {\it ArrDelay} can be well explained by those of {\it DepDelay} and {\it Distance}. As is shown in the result, the departure delay is a critical influential factor since it often directly leads to the arrival delay. The result also shows that given the same departure delay, a flight with a longer distance may have less arrival delay. When the distance is longer, the pilot may have more opportunities to adjust the flight and mitigate the delay.

The logistic model can be constructed through the {\it logit} function. For illustration, we construct a logistic regression model where the dependent variable is a binary variable transformed from {\it ArrDelay}. To explore the influence of the departure time on the arrival delay, we choose {\it DayOfWeek} (day for week) and {\it DepTime} (the real departure time) as independent variables. In the airline dataset, the values of {\it DepTime} are all recorded as numbers. For example, ``12:00'' is denoted as ``1200''. In this case, {\it DepTime} is binned into 4 categories: ``morning'' (700-1200), ``afternoon'' (1200-1800), ``evening'' (1800-2400) and ``midnight'' (0-700, used as the base group), respectively. For {\it DayOfWeek}, ``Sunday'' is treated as the base group. The construction of this logistic model is shown as follows.
\begin{lstlisting}
>>> mycodebook={'qlist': ['ArrDelay', 'DepTime', 'DayOfWeek','_INTERCEPT_']}
>>> mypathfile='2008.csv.shuffle'
>>> pm=cb.pump(type=2, pathfile=mypathfile, codebook=mycodebook)
>>> pm.subsize=10**5
>>> pm.seq=True
>>> def generateDepTimeBins(x):
        if x<700:
            return 'midnight'
        elif x<1200:
            return 'morning'
        elif x<1900:
            return 'afternoon'
        else:
            return 'evening'
>>> tk=cb.tank(pm)
>>> tk.app(lambda x: generateDepTimeBins(x), 'DepTime')
>>> tk.ady('DepTime', ['morning', 'afternoon', 'evening'])
>>> tk.ady('DayOfWeek', [1,2,3,4,5,6])
>>> tk.app(lambda x: 1 if x>0 else 0,'ArrDelay')
>>> md=cb.model(tk).logit('ArrDelay', niter=20, tv=True)

Time elapsed: 62.2 seconds with averaged subsample sizes 100000.0
Task accomplished:  100.0 % for a total of  20 random replications.
The out-of-sample AUC =  56.8 %

                   Estimate  StandErr    tStat  pValue
DayOfWeek_1          -0.005     0.537   -0.945  34.442
DayOfWeek_2          -0.077     0.539  -14.333   0.000
DayOfWeek_3          -0.076     0.538  -14.196   0.000
DayOfWeek_4           0.053     0.537    9.814   0.000
DayOfWeek_5           0.149     0.535   27.893   0.000
DayOfWeek_6          -0.143     0.567  -25.136   0.000
DepTime_afternoon     0.618     0.555  111.454   0.000
DepTime_evening       0.711     0.612  116.280   0.000
DepTime_morning       0.284     0.572   49.585   0.000
_INTERCEPT_          -0.770     0.629 -122.510   0.000

* Estimate: the ordinary least squares estimate for standardized x variables.
* StandErr: the standard error in %; t.Stat: the t-statistics = Estimate/Stand.Err.
* pValue: the p-value computed by normal approximation in %.
\end{lstlisting}
In this code, we generate a tank and adopt the function {\it app} to implement variable transformations. It is noteworthy that {\it app} is very flexible. It supports functions customized by users. In this case, we define a customized function {\it generateDepTimeBins(x)} to transform {\it DepTime} into a categorical variable with 4 levels. In addition, {\it CluBear} provides a function {\it ady} for the generation of dummy variables. By {\it ady}, we specify a categorical variable with the list of required levels. The left level is used as the base group. By setting {\it tv=True} in the function {\it logit}, the regression result will be shown in a table. If we set {\it tv=False} in the function {\it logit}, the result will be visualized with an informative bar plot, as is shown in Figure \ref{fig:lrpic}. The values of $T$ statistic for all coefficients are illustrated in the bar plot. According to the colors of the bars, users can distinguish significant variables (red) and insignificant variables (purple), e.g., {\it DayOfWeek\_1} in Figure \ref{fig:lrpic}. With respect to the day in a week, flights on Friday are most likely to be delayed. In terms of the departure time, delays are more likely to happen for flights departing in the afternoon or evening.
\begin{lstlisting}
>>> md=cb.model(tk).logit('ArrDelay', niter=20, tv=False)
\end{lstlisting}
\begin{figure}
	\centering
	\includegraphics[width=0.8\linewidth]{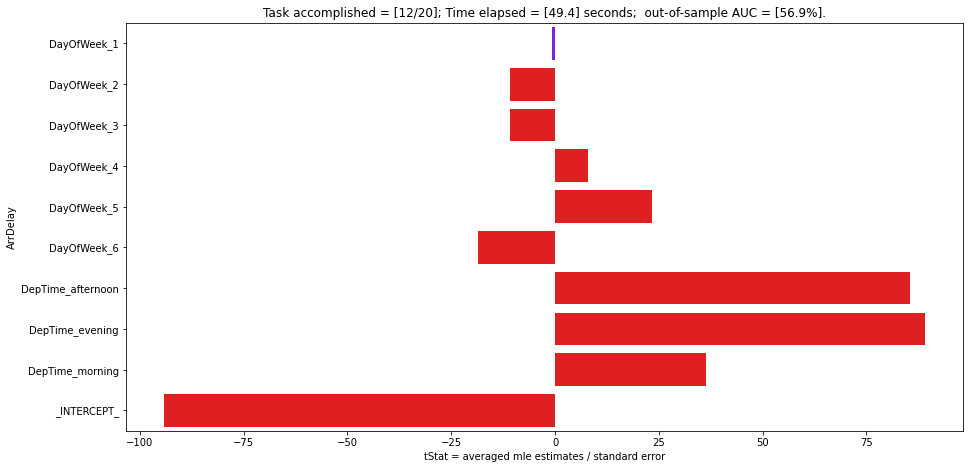}
	\caption{\label{fig:lrpic}
		The barplot of {\it tStat} for the logistic regression model}
\end{figure}

}

\section{Conclusion} \label{sec:conclusion}
In this article, we present the Python package {\it CluBear}, which supports the interactive statistical analysis for massive datasets. Through subsampling-based techniques and a number of built-in functions, {\it CluBear} can help users interactively analyze massive data with low-cost devices (e.g., a personal computer). These functions allow users to extract, clean, visualize and model data conveniently. Thus, with regard to the analysis of massive data, the package {\it CluBear} will contribute to a user-friendly and cost-effective solution for most general users. All the statistical analyses and graphical displays provided by {\it CluBear} are interactive in nature. Because subsamples are repeatedly generated, all the statistical tables or graphical displays are also updated dynamically, which creates a pleasant user experience. One can gain a quick understanding of experience by either directly trying {\it CluBear} or watching a video clip at https://www.youtube.com/watch?v=00Qu8Bq\_lXU. The current package is developed for Jupyter NoteBook environment only. We wish to be able to support more programming environments in the future.

\section*{Acknowledgements}
This work was supported by the National Natural Science Foundation of China (No. 12001102, 72301070), and "the Fundamental Research Funds for the Central Universities" in UIBE (NO. CXTD13-04, CXTD14-05, 22QD09).

\section*{Declarations}
We have read and understood the journal's policies, and we believe that neither the manuscript nor the study violates any of these. There are no conflicts of interest to declare.

\bibliographystyle{tfnlm}

\bibliography{prefs}

\end{document}